\documentclass[preprints,article,accept,moreauthors,color,dvi2pdf]{mdpi} 
\firstpage{1}
\pubvolume{xx}
\issuenum{1}
\articlenumber{5}
\pubyear{2019}
\copyrightyear{2019}
\usepackage{amssymb,amsmath,amsfonts,latexsym,graphicx,epsfig,fancybox}
\newcommand{\bea}{\begin{eqnarray}}
\newcommand{\ena}{\end{eqnarray}}
\newcommand{\eq}{\begin{eqnarray}}
\newcommand{\en}{\end{eqnarray}}
\newcommand{\nn}{\nonumber\\}

\newcommand{\Tr}{\mbox{\rm{tr}}}

\history{Received: 15 May 2019; Accepted: 10 June 2019; Published: 
13 June 2019}

\continuouspages{yes}

\Title{
\hfill {\small MITP/19-030 (Mainz)}\\[.5cm]
Novel ideas in nonleptonic decays of double heavy baryons} 

\Author{Thomas Gutsche $^{1}$, 
Mikhail A. Ivanov $^{2}$, 
J\"urgen G. K\"orner $^{3}$, 
Valery E. Lyubovitskij $^{1,4}$  
}

\address{
$^{1}$ \quad 
Institut f\"ur Theoretische Physik, Universit\"at T\"ubingen,
Kepler Center for Astro and Particle Physics, \\
\hspace*{.5cm} Auf der Morgenstelle 14, D-72076 T\"ubingen, Germany\\
$^{2}$ \quad 
Bogoliubov Laboratory of Theoretical Physics,
Joint Institute for Nuclear Research, 141980 Dubna, Russia\\
$^{3}$ \quad 
PRISMA$^+$ Cluster of Excellence, Institut f\"{u}r Physik,
Johannes Gutenberg-Universit\"{a}t, D-55099 Mainz, Germany\\
$^{4}$ \quad 
Departamento de F\'\i sica y Centro Cient\'\i fico
Tecnol\'ogico de Valpara\'\i so-CCTVal, \\ 
\hspace*{.5cm} Universidad T\'ecnica Federico Santa Mar\'\i a, 
Casilla 110-V, Valpara\'\i so, Chile 
}
\corres{Correspondence: valeri.lyubovitskij@uni-tuebingen.de} 

\abstract{The recent discovery of double charm baryon states by 
the LHCb Collaborarion and their high precision mass determination 
calls for a comprehensive analysis of the nonleptonic 
decays of double and single heavy baryons. Nonleptonic baryon decays play 
an important role in particle phenomenology since they allow to study the
interplay of long and short distance dynamics of the Standard Model (SM).
Further, they allow one to search for New Physics effects beyond the SM. We
review recent progress in experimental 
and theoretical studies of the nonleptonic decays of heavy baryons with a focus 
on double charm baryon states and their decays. In particular, 
we discuss new ideas proposed by the present authors to calculate the 
$W$--exchange matrix elements of the nonleptonic decays of double heavy
baryons. An important ingredient in our approach 
is the compositeness condition of Salam and Weinberg, and an effective
implementation of infrared confinement both of which allow one to describe 
the nonperturbative structure of baryons composed of light and heavy quarks. 
Further we discuss an {\it ab initio} calculational method for the treatment of 
the so-called $W$--exchange diagrams generated by $W^{\pm}$ boson exchange
between quarks. We found that the $W^{\pm}$--exchange contributions are not
suppressed in comparison 
with the tree-level (factorizing) diagrams and must be taken 
into account in the evaluation of matrix elements. Moreover, there are
decay processes such as 
the doubly Cabibbo-suppressed decay $\Xi_c^+ \to p \phi$ recently observed
by the LHCb Collaboration 
which is contributed to only by one single $W$--exchange diagram.} 


\begin{document}

{\bf Foreword}  

This paper is written in memory of Garry Efimov (1934–2015). 
For two of us (Mikhail Ivanov and Valery Lyubovitskij) 
Garry Efimov was a teacher and subsequently an important collaborator.
 He significantly contributed to the development of 
nonlocal quantum field theory and its relation to hadron structure. 
In particular, he showed that the nonlocal structure of hadron-quark 
interactions is important for a realistic description of the nonperturbative 
features of the strong interaction contributions at large distances. 
 Phenomena such as hadronization and confinement can be consistently 
implemented in field-theoretical approaches based on nonlocal 
phenomenological Lagrangians. 
Garry Efimov with his pupils and collaborators developed a series of 
relativistic quark models based on nonlocal Lagrangians. The covariant
constituent quark model (CCQM) developed by us and described in the present
paper is deeply rooted in his ground-breaking ideas on the use of nonlocal
quantum field theory in particle physics.

\section{Introduction}

The nonleptonic decays of light and especially heavy baryons 
play an important role in the phenomenology of particle interactions. The
nonleptonic decays can be used 
to determine some of the parameters of the Standard Model (SM)  
and to search for New Physics beyond the SM. The last decades have seen 
significant experimental progress in the study of nonleptonic decays of 
heavy baryons. The CDF, ATLAS, LHCb, Belle and BES Collaborations have measured 
various features of the nonleptonic decays of heavy baryons~\cite{PDG18}. 
In particular, there are now more precise results on 
the branching ratios of the two-body decays 
of charm baryons $\Lambda_c^+ \to p \phi, \Lambda \pi^+, 
\Sigma^+ \pi^0$~\cite{Ablikim:2015flg} and $\Xi_c^+ \to p \bar
K^\ast(892)^o$~\cite{Li:2019atu} , and bottom baryons 
$\Lambda_b^0 \to \Lambda J/\psi, 
\Lambda \psi(2S)$~\cite{Aad:2015msa,Abazov:2011wt}.  
Starting in 2005 a new era began in the studies of double charm baryons  
when the SELEX Collaboration reported on the observation of a state with the
quantum numbers of the spin 1/2 ground state $\Xi_{cc}^+$ baryon with a mass of 
$3518 \pm 3$ MeV~\cite{Ocherashvili:2004hi}.
This double charm baryon state was conjectured to be an isospin-$\frac{1}{2}$ baryon 
with quark content $(dcc)$ and to have an isospin partner 
$\Xi_{cc}^{++}$ with the quark structure $(ucc)$. 
However, other Collaborations ({\it BABAR}, Belle, LHCb~\cite{PDG18}) 
found no evidence for the $\Xi_{cc}^+$ nor the $\Xi_{cc}^{++}$ states 
in the conjectured mass region of $\sim 3500 MeV$. Recently the LHCb
Collaboration  discovered the double charm state
$\Xi_{cc}^{++}$~\cite{Aaij:2017ueg}-\cite{Aaij:2018gfl} 
in the invariant mass spectrum of the final state particles 
$(\Lambda_c^+\,K^-\,\pi^+\,\pi^+)$. The extracted mass of the $\Xi_{cc}^{++}$ 
state was given as $3621.40 \pm 0.72 \pm 0.27 \pm 0.14$ MeV and was
$\sim 100$ MeV heavier than the mass of the original SELEX double charm baryon
state $\Xi_{cc}^+$ which made it quite unlikely that the two states were
isospin partners.
On the other hand, the LHCb mass measurement was in agreement with  
theoretical mass value predictions for the double charm baryon states.
In particular, the central mass value of 
the LHCb result for the $\Xi_{cc}^{++}$ was very close the value 
3610 MeV and 3620 MeV predicted in~\cite{Korner:1992wi,Korner:1994nh} using
the one gluon 
exchange model of de Rujula, Georgi and Glashow~\cite{DeRujula:1975smg} 
and a relativistic quark-diquark potential model~\cite{Ebert:2002ig}, 
respectively. Fleck and Richard using a variety of models had also predicted
a mass value of $\sim 3600$~\cite{Fleck:1989mb} while Karliner et al.
found $M_{\Xi_{cc}}=3627 \pm 12$ MeV~\cite{Karliner:2014gca}.

\begin{figure}
\begin{center}
\epsfig{figure=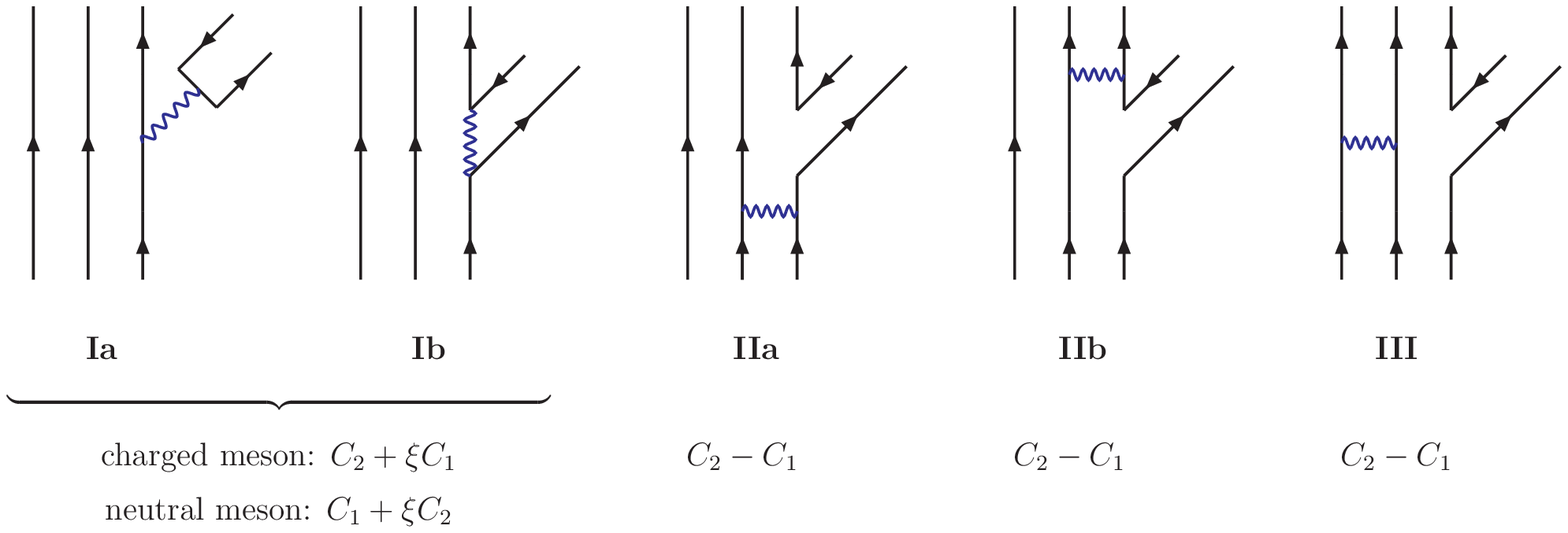,scale=.75}
\caption{Flavor-color topologies of nonleptonic two-body decays.}
\label{fig:NLWD}
\end{center}
\end{figure}

The new measurement of the LHCb Collaboration has stimulated much
theoretical activity concerning the structure of the nonleptonic decays of
double heavy baryons. 
The nonleptonic two-body baryon decays can be conveniently classified by 
the color-flavor quark level topologies with a single effective $W$-exchange
between
the constituent quarks participating in the decay process. The corresponding
set of topological quark diagrams are shown in Fig.~\ref{fig:NLWD}. They can
be divided into two subgroups: i) the reducible so-called
tree-diagrams Ia and Ib, and ii) the irreducible  $W$-exchange diagrams
IIa, IIb and III. Further details 
can be found in
Ref.~\cite{Korner:1978tc,Korner:1992wi,Korner:1994nh,Leibovich:2003tw}.
The tree-diagrams Ia and Ib can be factorized 
into the lepton decay of the emitted meson and a baryon-baryon
transition matrix element induced by the relevant weak currents.
The $W$--exchange diagrams IIa, IIb and III are more
difficult to evaluate from first principles. 

The two-body nonleptonic decays of the double charm baryon states
$\Xi_{cc}^{+,++}$ were
studied in Ref.~\cite{Sharma:2017txj} by using factorization for the
tree diagrams and a pole model for the $W$--exchange contributions.
The nonfactorizable $W$--exchange contributions
obtained in their pole model approach were found to be sizable and provide
an indication that the $W$--exchange contributions cannot be ignored. 
In Ref.~\cite{Dhir:2018twm} the same approach has been applied to
the nonleptonic decay modes of the double charm strange
$\Omega_{cc}^+$ baryon.  It was found that the branching ratios of some of the
nonleptonic $\Omega_{cc}^+$ decay modes obtain contributions solely from 
$W$-exchange diagrams, resulting in branching ratios
as large as $\sim{\cal{O}}(10^{-2})$.  The $W$-exchange 
contributions in double charm decays have been found to be comparable to the
factorizable contributions for most of the decay modes. In the paper by  
Zhang et al.~\cite{Zhang:2018llc} the $W$-exchange contributions have been
analyzed
for the nonleptonic transitions of the double heavy baryon states $\Xi_{bc}$
and $\Omega_{bc}$ using $SU(3)$ flavor symmetry.
The decay amplitudes for the various decay channels were parameterized
in terms of $SU(3)$ irreducible amplitudes resulting in
a number of relations between the decay widths.
In Ref.~\cite{Wang:2017mqp}
various nonleptonic decay modes
of double charm, bottom-charm, and double bottom baryons have been
analyzed using a light-front QCD approach for the calculation of
hadronic form factors induced by the $c \to d(s)$ and
$b \to u(c)$ quark-level transitions.
A detailed theoretical analysis of the production, lifetime and
semileptonic and nonleptonic decays of double heavy baryons
has been performed in
Refs.~\cite{Kiselev:2001fw,Berezhnoy:2018bde} by using nonrelativistic QCD.
The analysis resulted in a number of predictions for various exclusive
decay modes (nonleptonic modes
with $\pi$, $\rho$ mesons in the final state and semileptonic modes). 
The authors of Ref.~\cite{Jiang:2018oak} studied the nonleptonic decays of
double charm baryon into single charm baryons and light
vector mesons in a phenomenological approach
based on the factorization hypothesis to evaluate short-distance
effects, while long-distance effects were modelled as final-state
interactions and were estimated using a one-particle-exchange model.

Our group has been studying nonleptonic decays 
of both single and double heavy baryons for a number of
years~\cite{Ivanov:1997ra}-\cite{Gutsche:2018msz}. Use was made of 
a covariant constituent quark model (CCQM) based on
phenomenological nonlocal Lagrangians describing the coupling of single and
double heavy baryons to their constituent quarks. Common to all our studies 
is the use of the Weinberg-Salam compositeness condition formulated in
Refs.~\cite{Salam:1962ap}-\cite{Efimov:1993ei}. The compositeness condition
allows one to determine the coupling constant between the hadrons and their
constituent quarks. The compositeness condition
has been successfully applied to the description of a variety of hadrons and
exotic states such as tetraquarks and 
hadronic molecules
(see, e.g., Refs.~\cite{Efimov:1993ei}-\cite{Faessler:2007gv}). 
In the later versions of our covariant constituent quark model
we have implemented the infrared (IR) confinement of quarks by a cut in the
relevant scale integration at a value $\lambda_{\rm IR} = 181$ MeV. The
infrared cut-off was taken to be universal for all processes considered in
this approach. 

Let us review our results on the nonleptonic decays of heavy baryons
described in~\cite{Ivanov:1997ra}-\cite{Gutsche:2018msz} in more detail.
In a 1998 precursor to the present version of the CCQM model we presented
a comprehensive analysis of heavy-to-heavy and heavy-to-light nonleptonic
transitions involving spin-$\frac{1}{2}$
baryons~\cite{Ivanov:1997ra,Ivanov:1997hi}. Instead of using full dynamic
quark propagators we employed static approximations 
for the light quark $q=u,d,s$ propagators and leading-order contributions
for the heavy quark $Q=c,b$ propagators in the $1/m_{c/b}$ expansion,
i.e. in the heavy quark limit. We included the factorizing tree diagrams as well
as the nonfactorizing $W$--exchange contributions to the decay amplitudes.
We found that for heavy-to-light transitions $Q \to qud$ the total contribution
of the nonfactorizing diagrams amounted up to $\sim 60\%$
of the factorizing contributions in amplitude, and up to
$\sim 30\%$ for $b \to c\bar ud$ transitions. 
We calculated the rates and polarization parameters for various
nonleptonic modes of baryons and compared them to
data and existing theoretical predictions.

In Ref.~\cite{Gutsche:2013oea} we calculated the invariant and helicity
amplitudes 
for the nonleptonic decay $\Lambda_b \to \Lambda + J/\psi,\, \psi(2S)$ in the 
CCQM model. We discussed joint angular decay distributions 
in the cascade decay 
$\Lambda_b \to \Lambda(\to p \pi^-) + J/\psi, \,\psi(2S) (\to \ell^+ \ell^-)$ 
and calculated some of the asymmetry parameters characterizing the joint
angular 
decay distribution. We confirmed expectations from the naive quark model that 
the transitions into the $\lambda_\Lambda=1/2$ helicity states of the daughter 
baryon $\Lambda$ are strongly suppressed leading to a near maximal negative 
polarization of the $\Lambda$. For the same reason the azimuthal correlation 
between the two decay planes spanned by $(p \pi^-)$ and $(\ell^+ \ell^-)$ is 
negligibly small. We provided form factor results for the whole accessible
range of $q^2$--values. 
Our results were found to be close to lattice QCD results at minimum recoil and 
light-cone QCD sum rule results at maximum recoil. A new feature of our
analysis was that we included lepton mass effects 
in the calculation which allowed us also to describe the cascade decay 
$\Lambda_b \to \Lambda(\to p \pi^-) + \psi(2S)(\to \tau^+ \tau^-)$. 
Our prediction for the branching ratio 
$R(\Lambda_b) = \Gamma(\Lambda_b^0 \to \psi(2S) \Lambda^0)/
\Gamma(\Lambda_b^0 \to J/\psi \Lambda^0) = 0.8 \pm 0.1$ differed from 
the measured branching ratio 
$R(\Lambda_b) = 0.501 \pm 0.033 ({\rm stat}) \pm 0.019 ({\rm syst})$
by the ATLAS Collaboration~\cite{Aad:2015msa} by 2.8 standard deviations. 

In Ref.~\cite{Gutsche:2015lea} we presented 
a detailed analysis of the above branching ratio $R(\Lambda_b)$ using 
a model-independent framework for the heavy-to-light form factors in which
the values of the form factors grow when going from
$q^2=m^2(J/\psi)$ to $q^2=m^2(\psi(2S))$. Taking into account phase-space
suppression effects as well as the difference of the leptonic decay
constants $f_{J/\psi}$ and $f_{\psi(2S)}$ we obtained
$R(\Lambda_b) = 0.81$ in agreement with our previous result published in 
Ref.~\cite{Gutsche:2015lea}. The small $R(\Lambda_b)$ value measured by the
ATLAS Coll. was confirmed by the LHCb Coll. who found 
$R(\Lambda_b) = 0.513 \pm 0.023 ({\rm stat}) \pm 0.016 ({\rm syst}) \pm 0.011$ 
\cite{Aaij:2019dvk}. The small experimental rate ratio values are very
puzzling since they imply decreasing $\Lambda_b \to \Lambda$ form factors for increasing $q^2$--values
in the charmonium mass region which is very counter-intuitive and contradicts
all form factor models in the literature. 

In Ref.~\cite{Gutsche:2017wag} we again used the CCQM model
to calculate  invariant and helicity amplitudes for the transitions 
$\Lambda_b \to \Lambda^{(*)}(J^P) + J/\psi$ where the $\Lambda^{(*)}(J^P)$ 
are $\Lambda(s[ud])$-type ground and excited states with $J^P$ quantum numbers 
$J^P=\frac{1}{2}^\pm, \frac{3}{2}^\pm$. We found that the values of the helicity 
amplitudes for the $\Lambda^*(1520,\frac{3}{2}^-)$ 
and the $\Lambda^*(1890,\frac{1}{2}^-)$ are suppressed compared to those for 
the ground state $\Lambda(1116,\frac{1}{2}^+)$ and 
the excited state $\Lambda^*(1405,\frac{1}{2}^-)$. We emphasized that our
analysis is important for the identification of the hidden charm pentaquark
states $P^+_c$ which were discovered in the decay chain 
$\Lambda^0_b \to P^+_c (\to p J/\psi)+ K^-$ by the LHCb 
Collaboration~\cite{Aaij:2015tga} because the cascade decay chain 
$\Lambda^0_b \to \Lambda^*(\frac{3}{2}^\pm (\to pK^- )  + J/\psi$ 
involves the same final state. 

In Ref.~\cite{Gutsche:2018utw} we have made a comprehensive analysis of 
heavy-to-heavy and heavy-to-light nonleptonic
heavy baryon two-body decays and have identified those decays
that proceed solely via $W$-boson emission, i.e. via the
tree graph contribution. Some sample decays are
$\Omega_{b}^{-}\to\Omega_{c}^{(*)0}\rho^{-}(\pi^{-}),\,                         
\Omega_{b}^{-}\to\Omega^{-}J/\psi(\eta_{c}),\,                                  
\Xi_{b}^{0,-}\to\Xi^{0,-}J/\psi(\eta_{c}),\,                                    
\Lambda_{b}\to \Lambda J/\psi(\eta_{c}),\,                                      
\Lambda_{b}\to \Lambda_{c} D_{s}^{(\ast)},\,                                    
\Omega_{c}^{0}\to\Omega^{-}\rho^{+}(\pi^{+})$
and $\Lambda_c \to  p \phi$.
We made use of the CCQM 
to calculate the tree graph contributions to these decays.
We calculated rates, branching fractions and, for some of these decays,
decay asymmetry parameters taking into account lepton mass effects. 
We compared our results to experimental results
and the results of other theoretical approaches when they were available.
Our main focus was on decays to final states with a lepton pair because of
their clean experimental signature.
For these decays we discussed two-fold polar angle decay distributions such
as in the cascade decay
$\Omega_{b}^{-}\to\Omega^{-}(\to \Xi\pi,\Lambda K^{-})+J/\psi(\to 
\ell^{+}\ell^{-})$. 

In Ref.~\cite{Gutsche:2017hux} we interpreted the new double charm baryon
state found by the LHCb Collaboration in the invariant mass distribution of
the set of final state particles 
$(\Lambda_c^+\,K^-\,\pi^+\,\pi^+)$ as being at the origin of the decay chain
$\Xi_{cc}^{++} \to \Sigma_c^{++} (\to \Lambda_c^+ \pi^+)
+ \bar K^{*0} (\to K^-  \pi^+)$. 
The nonleptonic decay $\Xi_{cc}^{++} \to \Sigma_c^{++} + \bar K^{*0}$ belongs 
to a class of decays where the quark flavor composition is such that the
decay proceeds solely via the factorizing tree-graph contribution precluding a
contamination from $W$--exchange. 
We used the CCQM model to calculate the four helicity amplitudes that 
describe the dynamics of the transition $\Xi_{cc}^{++} \to \Sigma_c^{++}$ 
induced by the effective $(c \to u)$ current. We then 
calculated the rate of the decay as well as the polarization of the 
$\Sigma_c^{++}$ and $\Lambda_c^+$ daughter baryons and the 
longitudinal/transverse composition of the $\bar K^{*0}$. 
We estimated the decay $\Xi_{cc}^{++} \to \Sigma_c^{++} \bar K^{*0}$
to have a branching rate of 
$B(\Xi_{cc}^{++} \to \Sigma_c^{++} \bar K^{*0}) \sim 10.5 \%$. 
As a byproduct of our investigation we have also analyzed the decay 
$\Xi_{cc}^{++} \to \Sigma_c^{++} \bar K^{0}$ for which we find 
a branching ratio of 
$B(\Xi_{cc}^{++} \to \Sigma_c^{++} \bar K^0) \sim 2.5 \%$.

In Ref.~\cite{Gutsche:2018msz} we performed an {\it ab initio} calculation of
the $W$-exchange contribution to nonleptonic decays of double charm
baryons  $\Xi_{cc}^{++}$ and $\Omega_{cc}^+$ based on their three-quark structure.
Preliminary approximate
calculations~\cite{Korner:1994nh,Ivanov:1997ra,Ivanov:1997hi}
had indicated that the $W$-exchange contribution are not negligible.
Prior to the above paper~\cite{Gutsche:2018msz} there
existed no first principles calculation of the $W$-exchange contribution. 
Again we used the CCQM model to calculate the tree graph
contribution as well as the $W$-exchange contribution induced by diagram IIb.
We calculated helicity amplitudes and determined the relative strengths
of the two contributions to the helicity amplitudes. We found that the
contribution of the $W$-exchange diagrams are suppressed in comparison
with the tree-level contributions for the decay modes involving
the flavor-symmetric final state charm baryon $\Xi^{\prime +}_c$.
For the decay modes involving
the flavor-antisymmetric final state charm baryon $\Xi^{+}_c$ the $W$--exchange
contributions are not
suppressed and even dominate over the tree-level contributions. We found that
the $W$--exchange and tree diagram contributions are destructive for the
decays into the $\Xi^{+}_c$ state.
Finally, we compared the calculated decay widths with those from other
theoretical approaches when they were available.

The main goal of our present paper is to provide a detailed description of
our novel 
idea to calculate $W$-exchange diagrams occurring in nonleptonic decays 
of baryons (both light and heavy) in the context of our CCQM model which is 
based on the use of phenomenological nonlocal hadron-quark interaction
Lagrangians. 

The paper is structured as follows. In Sec.~II we briefly review some basic
notions of our CCQM model which are 
i) the underlying phenomenological nonlocal Lagrangians, ii) the implementation
of hadronization and quark confinement, iii) the calculation of matrix
elements, iv) the adjustment of the model parameters. We recount some
important physical applications to the description
of hadrons and their decays in our CCQM approach. 
In Sec.~III we discuss our novel method of how to evaluate the $W$-exchange 
diagrams which involve the calculation of three-loop quark diagrams. We also
discuss applications 
of our method to the nonleptonic decays of double charm baryons. 
In Sec.~IV we present our numerical results for some specific nonleptonic decay
modes of the double charm baryons 
$\Omega^+_{cc}\to\Xi^{\prime\,+}_{c}(\Xi^{+}_{c}) + \bar K^0(\bar K^{\ast\,0})$ and 
  $\Xi^{++}_{cc}\to\Xi^{\prime\,+}_{c}(\Xi^{+}_{c}) + \pi^+(\rho^+)$. 
Finally, in Sec.~V, we summarize our results and present prospects 
for future studies. 

\section{Covariant Constituent Quark Model}

In this section we describe the main features of the covariant constituent quark
model (CCQM)~\cite{Branz:2009cd,Gutsche:2015rrt,Dubnicka:2010kz} which will
be used as a theoretical tool to address the hadron structure input in
the study of nonleptonic decays of heavy baryons. 
The CCQM is a universal, truly relativistic, and a manifestly
Lorentz covariant quark model for the description of hadrons as bound states
of constituent quarks and of exotic states (hadronic molecules, tetraquarks,
pentaquarks, hybrids, etc.~\cite{Dubnicka:2010kz,Faessler:2007gv}. 
Our approach allows one to study bound states with an arbitrary number of
constituents 
and with arbitrary quantum numbers (spin-parity, isospin, flavor content, etc.)
The CCQM is based on a phenomenological,
nonlocal relativistic Lagrangian describing the coupling of a hadron to its
constituents. As an example we write down the Lagrangian describing the
coupling of a baryon
$B(q_1q_2q_3)$ to its three  constituent quarks $q_1$, $q_2$, and $q_3$.
The Lagrangian has the form
\eq\label{Lagr}
{\cal L}_B(x) &=& g_B \bar B(x) J_B(x) \, + \, {\rm H.c.}\,, \nonumber\\
J_B(x)&=& \int\!\! dx_1 \!\! \int\!\! dx_2 \!\! \int\!\! dx_3 \,
F_B(x;x_1,x_2,x_3) \,
\varepsilon^{a_1a_2a_3}\,\Gamma_1\, q^{a_1}(x_1)\,
\left(q^{a_2}(x_2) \,C\Gamma_2 \, q^{a_3}(x_3)\right)\,,
\nonumber\\
F_B(x;x_1,x_2,x_3) &=& \delta^{(4)}\Big(x-\sum\limits_{i=1}^3 w_i x_i\Big)
\Phi_B\Big(\sum\limits_{i<j}(x_i-x_j)^2\Big) \,,
\label{eq:cur}
\en
where $J_B$ is an interpolating three-quark current with the quantum numbers
of the baryon state $B$. Further $w_i=m_i/(\sum\limits_{j=1}^3 m_j)$ where
$m_i$ is the quark mass associated with the space-time point $x_i$.
$\Gamma_1$ and $\Gamma_2$ are Dirac matrix strings.
$F_B$ is the Bethe-Salpeter kernel specifying the
coupling of the baryon to the constituent quarks.
The vertex form factor (or correlation function) $\Phi_B$ is a
process-independent function which depends on the relative (Jacobi)
coordinates and encodes
the information concerning the distribution of the constituents in the
bound state. A basic requirement for the choice of an explicit
form of the correlation function $\Phi_B$ is that its Fourier
transform vanishes sufficiently fast in the ultraviolet region of
Euclidean space to render the Feynman diagrams ultraviolet finite.

Note that Eq.~(\ref{eq:cur}) is not the only possible choice that is
compatible with Lorentz invariance. Any combination as
$ \sum_{i<j} (x_i - x_j)^2/r^2_{ij}$ preserves both translational and
Lorentz invariance. Generally speaking,
the correlation $r_{cd}$ should be different from $r_{cc}$.
However, it will increase the number of adjustable
parameters, i.e. one needs to keep different values for
$r_{cd}$, $r_{cs}$ and $r_{cc}$. But there is no much data even for
the single-charmed baryon. The double-charmed baryon was discovered
very recently by LHCb-collaboration. For that reasons, as a first
approximation we equate the size parameter of double charm baryons with
that of single charm baryons, i.e. we take
$\Lambda_{cc}=\Lambda_{c}=0. 8675$~GeV where we
adopt the value of $\Lambda_{c}$ from~\cite{Gutsche:2015rrt}.

For simplicity and calculational advantages we mostly adopt a Gaussian form
for the correlation functions, i.e. we write
\bea
\Phi_B\Big(\sum\limits_{i<j}(x_i-x_j)^2\Big) &=&
\int\!\frac{dq_1}{(2\pi)^4}\int\!\frac{dq_2}{(2\pi)^4}
\widetilde\Phi_B\Big(-\tfrac12 (q_1+q_2)^2-\tfrac16 (q_1-q_2)^2\Big)\,,
\nn
\widetilde\Phi_B\Big(-\vec\Omega^2\Big) &=&
\exp\left(\vec\Omega^2/\Lambda_B^2\right)
\label{eq:vertex}
\ena
where $\Lambda_B$ is the size parameter for a given
baryon with values of the order of 1 GeV. The size parameter 
represents the extension of the distribution of the constituent quarks in the
given baryon. The values of the size parameters $\Lambda_B$
are fixed using data on fundamental properties of mesons and baryons such as
leptonic decay constants, magnetic moments and radii. We emphasize
that the Minkowskian momentum variable $p^2$ turns into the
Euclidean form $-p_E^2$ needed for the appropriate fall-off behavior of the
correlation function $\widetilde\Phi_B$ in the Euclidean region.
Any choice for the correlation function $\Phi_B$ is acceptable as long
as it falls off sufficiently fast in the ultraviolet region of
Euclidean space.

For given values of the size parameters $\Lambda_B$ the coupling constant
$g_B$ is determined by the compositeness condition suggested by
Weinberg and Salam~\cite{Salam:1962ap,Weinberg:1962hj} and extensively
used in our approach. The compositeness condition implies that the
renormalization constant of the hadron wave function is set equal to zero
$Z_B = 1 - \Sigma^\prime_B(m_B) = 0$,
where $\Sigma^\prime_B$ is the on-shell derivative of the
hadron mass function $\Sigma_B$ with respect to its momentum. The
compositeness condition can be seen to provide for the correct charge
normalization of a charged bound state.

\begin{figure}[H]
\begin{center}
\epsfig{figure=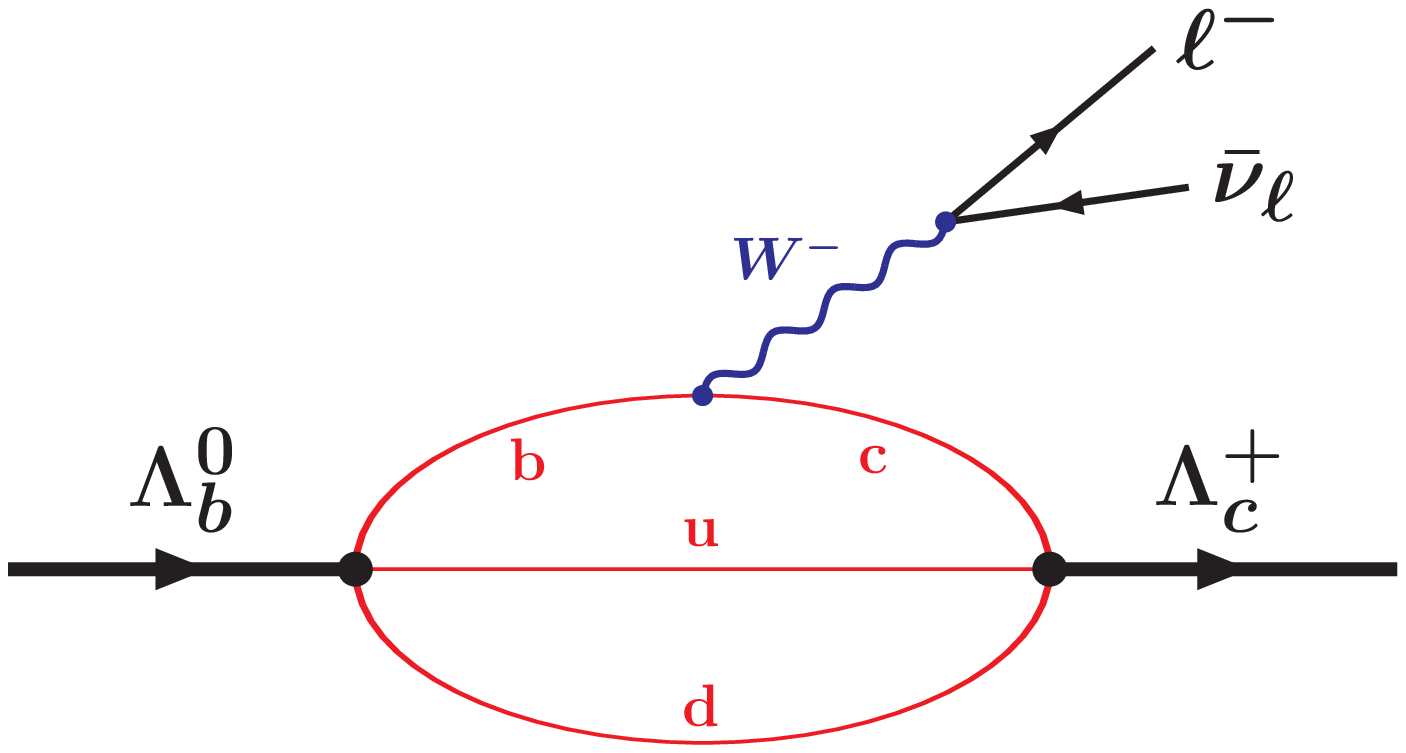,scale=.5}
\caption{Feynman diagram describing the semileptonic decay
$\Lambda_b^0 \to \Lambda_c^+ \ell^- \bar\nu_\ell$ decay.}
\label{fig:LbLcenu} 
\end{center}
\end{figure}

Matrix elements of hadronic processes are computed in the
$S$-matrix formalism, where the $\hat{S}$-operator is constructed
from the interaction Lagrangian ${\cal L}_{\rm int}$
\eq
\hat{S} = \hat{T} \exp\Big[i \int d^4x \, {\cal L}_{\rm int}(x)\Big] \,,
\en
and where $\hat{T}$ is the time-ordering operator.
The Lagrangian ${\cal L}_{\rm int}$ includes two parts i) the model
interaction Lagrangian which describes the coupling of hadrons to their
constituent quarks (see Eq.~(\ref{Lagr}))
which embodies the nonperturbative strong interaction effects at large
distances and II) the
electroweak part together with strong short-distance effects at the order
of accuracy we are working in as given by the Standard Model (SM).
As an illustration we consider the semileptonic decay 
$\Lambda_b^0 \to \Lambda_c^+ \ell^- \bar\nu_\ell$, 
which is described by the two-loop Feynman diagram
shown in~Fig.~\ref{fig:LbLcenu}.

The semileptonic weak 
Lagrangian describing the $b \to c \ell^- \bar\nu_\ell$ transition is given by 
\eq 
{\cal L}_{\rm sl} = \frac{G_F}{\sqrt{2}} \, V_{cb} \, 
(\bar c O_\mu b) \, (\bar e O^\mu \nu_e)\,, \quad O_\mu = \gamma_\mu (1-\gamma_5)
\,,    
\en 
where $G_F = 1.16638 \cdot 10^{-5}$ GeV$^{-2}$ is the Fermi constant, 
$V_{cb} = 0.0406$ the Cabibbo-Kobayashi-Maskawa  matrix element. 

The diagrams appearing in our approach can be viewed as ordinary
Feynman diagrams in which the point-like particle-quark vertex functions are
replaced by nonlocal vertex functions. Since the vertex functions enter as a
building block in any given Feynman diagram it is clear that the nonlocal
vertex functions are universal and not process dependent. As in ordinary
Feynman diagrams one has free quark propagators
$S_q(k) = 1/(m_q - \not\!\!k )$ in terms of the constituent quark masses
$m_q$ with $q=u,d,s,c,b$. 
In particular, the invariant matrix element
corresponding to the diagram in Fig.~\ref{fig:LbLcenu} 
is written as
\bea
{\cal M}  &=&
\frac{G_F}{\sqrt{2}} \, V_{cb} \, g_{\Lambda_b} \, g_{\Lambda_c}
\int\! \frac{d^4k_1}{(2\pi)^4i}\! \int\! \frac{d^4k_2}{(2\pi)^4i} \,
\widetilde\Phi_{\Lambda_b}\Big(-\vec\Omega^2_{\,\rm in}\Big) \,
\widetilde\Phi_{\Lambda_c}\Big(-\vec\Omega^2_{\,\rm out}\Big)
\nn
&\times&
\bar u_{\Lambda_c}(p') \, \Gamma_{\Lambda_c}^f \,  S_c(k_1 + p') \, \Gamma_i
\, S_b(k_1 + p) \, \Gamma_{\Lambda_b}^f \, u_{\Lambda_b}(p) \,
{\rm Tr}\Big[ 
\Gamma_{\Lambda_c}^i S_u(k_1+k_2)
\Gamma_{\Lambda_b}^i S_d(k_2) \Big] \,.
\label{M_sl}
\ena
$\Gamma_i$ is the set of Dirac matrices defining the $b \to c$ transition,
$\Gamma_{\Lambda_c}^i$,
$\Gamma_{\Lambda_c}^f$ and
$\Gamma_{\Lambda_b}^i$,
$\Gamma_{\Lambda_b}^f$ are the pair of Dirac matrices
defining the spin structure of
the $\Lambda_c$ and $\Lambda_b$ baryon, respectively,
$g_{\Lambda_c}$ and $g_{\Lambda_b}$ are the coupling constants of
$\Lambda_c$ and $\Lambda_b$ baryons with their constituent quarks. 
$\widetilde\Phi_{\Lambda_c}$ and $\widetilde\Phi_{\Lambda_b}$ are the Fourier
transforms of the  correlation functions taking into account the distribution
of quarks in the  $\Lambda_c$ and $\Lambda_b$ baryons. The long-distance
strong interaction effects are encoded in the correlation functions.
The constituent quark masses $m_u = m_d = 0.241$ MeV,
$m_s = 0.428$~GeV, $m_c = 1.672$ GeV and $m_b = 5.046$ GeV are taken to be
universal for all processes considered in our formalism. Their values have been
determined by a fit to a multitude of processes.
The calculated matrix elements are expanded in terms of a set of relativistic
form factors dictated by Lorentz and gauge invariance.
It should be quite clear that the evaluation of the form factors is
technically quite
involved since it necessitates the calculation of two-loop  
Feynman diagrams with a complex spin structure resulting from
the quark propagators and the vertex functions. The spin structure of the
diagrams leads to a number
of tensor integrals. To tackle this task we have automated the calculation
in the form of FORM~\cite{Vermaseren:2000nd} and FORTRAN packages written
for this purpose.

We emphasize that the CCQM model described here is a truly
frame-independent field-theoretical quark model in contrast to other
constituent quark models which are often quantum-mechanical with built-in
relativistic elements. Also note that the HQET relations for the form factors
can be recovered
in the CCQM approach by using a $1/m_Q$ expansion for the heavy quark
propagator. One of the advantages of the CCQM model is that
it allows to calculate transition form factors in the full accessible
range of momentum transfers. Heavy quark symmetry, for example,
is expected to be reliable only close to zero recoil.

The local form of the quark propagators used in Eq.~(\ref{M_sl}) can lead
to unwanted singularities corresponding to free quark production in
transition amplitudes, i.e. when $\sum m_{q_i} \le m_B$. In a further
refinement of earlier versions of our CCQM model
we introduced effective infrared confinement by introducing
a universal and finite scale variable, which, in the first version of
the CCQM model without confinement,
tends to infinity. The introduction of such an infrared cutoff removes
all physical quark thresholds when $\sum m_{q_i} \le m_B$. The CCQM model
could thus be extended to cases where the hadron mass
exceeded the
sum of its constituents masses. The formalism was successfully 
applied to the decays of light and heavy mesons and baryons 
(see details in Refs.~\cite{Branz:2009cd}). 

In our papers the CCQM calculation of dynamical
quantities in terms of e.g. helicity amplitudes was always accompanied by a
model-independent derivation of the spin-kinematics of a given process.
The ensuing angular decay distributions have been quite useful in the
analysis of data and have been used by other
theoretical and experimental groups (see e.g. discussion in
Ref.~\cite{Meinel:2014wua}).
In Table~\ref{tab1} we show some of our recent results on weak
(semileptonic, radiative,
and nonleptonic) decays of single and double heavy baryons and compare them
to available data~\cite{PDG18}.

\begin{table}
\caption{Branching ratios of weak decays of single and
double heavy baryons (in \% )}
\label{tab1}
\def\arraystretch{1.2}
\begin{center}
    \begin{tabular}{|c|c|c|}
      \hline
Mode & Our results 
     & Data~\cite{PDG18} \\
\hline
$\Lambda_c \to \Lambda e^+\nu_e$
& $2.0$ & $2.1 \pm 0.6$ \\
\hline
$\Lambda_c \to \Lambda \mu^+\nu_\mu$
& $2.0$ & $2.0 \pm 0.7$ \\
\hline
$\Lambda_c \to p \phi$ & 14.0 & $10.8 \pm 1.4$ \\
\hline
$\Lambda_b \to \Lambda_c e^-\bar\nu_e$
& $6.6$ & $6.2^{+1.4}_{-1.3}$ \\
\hline
$\Lambda_b \to \Lambda \mu^+\mu^-$
& 1.0 $\times$ 10$^{-4}$ &  (1.08 $\pm$ 0.28)
$\times$ 10$^{-4}$ \\
\hline
$\Lambda_b \to  \Lambda \gamma$ & 0.4 $\times$ 10$^{-3}$
&  $<$ 0.13 \\
\hline
$\Lambda_b^0 \to \Lambda_c^+\,D_s^-$ &  147.8 & 110 $\pm$ 10 \\
\hline
$\Lambda_b^0 \to \Lambda^0 J/\psi$ & 8.3 & 8.3 $\pm$ 1.1 \\
\hline
$\Xi_{cc}^{++} \to \Sigma_{c}^{++} \bar K^{*0}$ & 5.4 & \\
\hline
$\Xi_{cc}^{++} \to \Sigma_{c}^{++} \bar K^{0}$ & 1.3 & \\
\hline
$\Xi_{cc}^{++} \to \Xi_{c}^{' +} \bar \rho^{+}$ & 16.7 & \\
\hline
$\Xi_{cc}^{++} \to \Xi_{c}^{' +} \bar \pi^{+}$ & 0.3 & \\
\hline
$\Xi_{cc}^{++} \to \Xi_{c}^{+} \bar \rho^{+}$ & 0.2 & \\
\hline
$\Xi_{cc}^{++} \to \Xi_{c}^{+} \bar \pi^{+}$ & 0.1 & \\
\hline
\end{tabular}
\end{center}
\end{table}

\section{Nonleptonic decays of heavy baryons: Evaluation of the $W$-exchange diagrams}

In this section we describe a novel method of how to evaluate the $W$-exchange
contributions to the matrix elements of the nonleptonic decays of heavy
baryons. 
In Table~\ref{tab2} we collect the quantum numbers of single
and double charm baryons and define the interpolating currents that describe
their nonlocal vertex structures.

\begin{table}
\begin{center}
  \caption{Quantum numbers and interpolating currents of single and
    double charm baryons}
\label{tab2}
\def\arraystretch{1.1}
\begin{tabular}{|l|c|c|c|}
\hline
Baryon\qquad &\quad $J^P$ \quad & \quad Interpolating current \qquad &
\quad Mass (MeV)
\\
\hline
$\Xi_{cc}^{++}$\quad \quad \quad & $\frac12^+$ &
$\varepsilon_{abc}\,\gamma^\mu\gamma_5 \, u^a (c^b C\gamma_\mu c^c)$  & 3620.6\\
$\Omega_{cc}^{+}$\quad \quad \quad & $\frac12^+$ &
$\varepsilon_{abc}\,\gamma^\mu\gamma_5 \, s^a (c^b C\gamma_\mu c^c)$ & 3710.0\\
\hline
$\Lambda_c^{+}$ & $\frac12^+$ &
$\varepsilon^{abc}\, c^a (u^bC\gamma_5 d^c$) & 2286.46 \\
$\Sigma_c^{++}$ & $\frac12^+$ &
$\varepsilon^{abc}\,\gamma^\mu\gamma_5 \, c^a (u^bC\gamma_\mu u^c$) & 2453.97 \\
$\Sigma_c^{+}$ & $\frac12^+$ &
$\varepsilon^{abc}\,\gamma^\mu\gamma_5 \, c^a (u^bC\gamma_\mu d^c$) & 2452.9 \\
$\Sigma_c^{0}$ & $\frac12^+$ &
$\varepsilon^{abc}\,\gamma^\mu\gamma_5 \, c^a (d^bC\gamma_\mu d^c$) & 2453.75 \\
$\Xi_{c}^{'+}$  & $\frac12^+$ &
$\varepsilon^{abc}\,\gamma^\mu\gamma_5 \, c^a (u^bC\gamma_\mu s^c$) & 2577.4\\
$\Xi_{c}^{'0}$  & $\frac12^+$ &
$\varepsilon^{abc}\,\gamma^\mu\gamma_5 \, c^a (d^bC\gamma_\mu s^c$) & 2577.9 \\
$\Xi_c^+$ & $\frac12^+$ & $\varepsilon_{abc}\, c^a (u^b C\gamma_5 s^c) $ & 2467.93\\
$\Xi_c^0$ & $\frac12^+$ & $\varepsilon_{abc}\, c^a (d^b C\gamma_5 s^c) $ & 2470.85\\
$\Omega_c^{0}$ & $\frac12^+$ &
$\varepsilon^{abc}\,\gamma^\mu\gamma_5 \, c^a (s^bC\gamma_\mu s^c$) & 2695.2 \\
$\Sigma_c^{*++}$ & $\frac32^+$ &
$\varepsilon^{abc}\,c^a (u^bC\gamma_\mu u^c$) & 2518.41\\
$\Sigma_c^{*+}$ & $\frac32^+$ &
$\varepsilon^{abc}\,c^a (u^bC\gamma_\mu d^c$) & 2517.5\\
$\Sigma_c^{*0}$ & $\frac32^+$ &
$\varepsilon^{abc}\,c^a (d^bC\gamma_\mu d^c$) & 2518.481\\
$\Xi_c^{*+}$ & $\frac32^+$ &
$\varepsilon^{abc}\,c^a (u^bC\gamma_\mu s^c$) & 2645.9\\
$\Xi_c^{*0}$ & $\frac32^+$ &
$\varepsilon^{abc}\,c^a (d^bC\gamma_\mu s^c$) & 2645.9\\
$\Omega_c^{*0}$ & $\frac32^+$ &
$\varepsilon^{abc}\,c^a (s^bC\gamma_\mu s^c$) & 2765.9\\
\hline
\end{tabular}
\end{center}
\end{table}

As has been discussed before, nonleptonic baryonic two-body decays can be
characterized by five types of topological color-flavor diagrams 
contributing to them (see Fig.~\ref{fig:NLWD}).
As an example we take the Cabibbo-favored nonleptonic decay modes of the
spin-1/2 ground state double charm baryon states
$\Xi_{cc}^{++},\,\Xi_{cc}^{+},\,\Omega_{cc}^+$. In Table~\ref{tab3} we record
which of the tree-level and $W$-exchange graphs contribute to a given single
decay mode. It is noteworthy that eight of the fifteen flavor decay modes
are contributed to only by $W$-exchange graphs.

\begin{table}
\caption{Classification of the Cabibbo favored
nonleptonic two-body decays of the ground state double charm baryons}
\label{tab3}

\begin{center}
\def\arraystretch{.95}
\begin{tabular}{|lccccc|}
\hline
\phantom{$1/2^+ \to 1/2^+ +0^-$}\qquad& ${\rm I_a}$ & ${\rm I_b}$ &
${\rm II_a}$ & ${\rm II_b}$ & III \\
\hline
      $\Xi_{cc}^{++} \to \Sigma_{c}^{(*)\,++} + \bar K^{(*)\,0}$\quad \quad \quad
      &-
      \quad \quad&$\surd$ \quad \quad& $-$\quad \quad& $-$ \quad \quad&$-$
      \\[0.5ex]
      $\Xi_{cc}^{++} \to \Xi_{c}^{(\prime,* )\,+} + \pi^+(\rho^+)$\quad \quad \quad
      &$\surd$
      \quad \quad&$-$ \quad \quad& $-$\quad \quad& $\surd$\quad \quad&$-$
      \\[0.5ex]
      $\Xi_{cc}^{++} \to \Sigma^{(*)\,+} + D^{(*)+}$\quad \quad \quad &$-$
      \quad \quad &$-$ \quad \quad & $-$\quad \quad& $\surd$\quad \quad&$-$
      \\[0.5ex]
      $\Xi^+_{cc} \to \Xi_{c}^{(\prime ,*)\,0}+ \pi^+(\rho^+)$ \quad \quad \quad &
      $\surd$\quad \quad &$-$\quad \quad &$\surd $\quad \quad &$-$\quad \quad &$-$
 \\[0.5ex]
$\Xi_{cc}^{+} \to \Lambda_c^{+}(\Sigma_c^{(*)+})                                                 
 +\bar K^{(*)0}$ &$-$ &$\surd$&$\surd$ &
$-$ &$-$\\[0.5ex]
$\Xi_{cc}^{+} \to \Sigma_c^{(*)++}                                                               
 + K^{(*)-}$ &$-$ &$-$ &
 $\surd$ &$-$&$-$\\[0.5ex]
 $\Xi^+_{cc} \to \Xi_c^{(\prime ,*)\,+}+ \pi^0(\rho^0)$
&$-$ &$-$ & $\surd$ &$\surd$&$-$\\[0.5ex]
 $\Xi^+_{cc} \to \Xi_c^{(\prime ,*)\,+}+ \eta(\eta^\prime)$
 &$-$ &$-$ &$\surd$ &$\surd$&$-$\\[0.5ex]
 $\Xi^+_{cc} \to \Omega_c^{(*)\,0} + K^{(*)\,+}$
&$-$ &$-$ &$\surd$ &$-$&$-$\\[0.5ex]
 $\Xi_{cc}^{+} \to \Lambda^{0}(\Sigma^{(*)0})                                                    
 + D^{(*)+}$ &$-$& $-$ &$-$ &$\surd$ &$\surd$
 \\[0.5ex]
 $\Xi_{cc}^{+} \to \Sigma^{(*)+}+ D^{(*)0}$
 &$-$ &$-$ &$-$& $-$ &$\surd$\\[0.5ex]
 $\Xi_{cc}^{+} \to \Xi^{(*)0}+ D_s^{(*)+}$
 &$-$ &$-$ &$-$& $-$ &$\surd$\\[0.5ex]
 $\Omega_{cc}^{+} \to \Xi_{c}^{(\prime ,*)\,+}                                                   
 +\bar {K}^{(*)0}$&$-$ &$\surd$ &$-$ &$\surd$ &$-$ \\[0.5ex]
 $\Omega_{cc}^{+} \to \Xi^{0\,(\prime ,*)}                                                       
 + D^{(*)+}$ &$-$ &$-$ &$-$ &$\surd$ &$-$ \\[0.5ex]
 $\Omega_{cc}^{+} \to \Omega_c^{0\,(*)}                                                          
 + \pi^+(\rho^+)$ &$\surd$ &$-$ &$-$ &$-$ &$-$ \\[0.5ex]
\hline
\end{tabular}
\end{center}
\end{table} 

The Cabibbo-favored quark level nonleptonic transitions $\bar s c\to \bar u d$
are governed by the effective Hamiltonian
\eq
\mathcal {H}_{\rm eff} & = &
- g_{\rm eff} \,
\left( C_1\,\mathcal{Q}_1 + C_2\,\mathcal{Q}_2\right)\, + \, {\rm H.c.},
\nonumber\\
\mathcal{Q}_1 &=&  (\bar s_a O_L c_b)(\bar u_b O_L d_a)
= (\bar s_a O_L d_a)(\bar u_b O_L c_b),
\nonumber\\
\mathcal{Q}_2 &=&  (\bar s_a O_L c_a)(\bar u_b O_L d_b)
= (\bar s_a O_L d_b)(\bar u_b O_L c_a),
\label{eq:eff-Ham}
\en
where we use the notation
$g_{\rm eff} = \frac{G_F}{\sqrt{2}} V_{cs} V^\dagger_{ud}$ and
$O^\mu_{L}=\gamma^\mu(1 - \gamma_5)$
for the weak matrices with left chirality.

The color-flavor factor of the tree
diagrams Ia and Ib depend on whether the emitted meson is charged or neutral.
For charged emission the color-flavor factor is given by the
combination of the Wilson coefficients $(C_2 + \xi C_1)$,
where $\xi=1/N_c$ and $N_c$ is the number of colors,
while for neutral emission the color-flavor factor
reads $(C_1 + \xi C_2)$. The $W$--exchange diagrams are more
difficult to evaluate and will be the subject of the following
paragraphs.

The nonlocal version of the interpolating currents shown in
Table~\ref{tab3} reads 
\bea
J_{B_{cc}}(x)  &=& \int\!\! dx_1 \!\! \int\!\! dx_2 \!\! \int\!\! dx_3 \,
F_{B_{cc}}(x;x_1,x_2,x_3) \,
\varepsilon_{a_1a_2a_3}\,\gamma^\mu\gamma_5\, q_{a_1}(x_1)\,
\left(c_{a_2}(x_2) \,C\gamma_\mu \, c_{a_3}(x_3)\right)\,,
\nn
J_{B_c}(x)  &=& \int\!\! dx_1 \!\! \int\!\! dx_2 \!\! \int\!\! dx_3 \,
F_{B_c}(x;x_1,x_2,x_3) \,
\varepsilon_{a_1a_2a_3}\,\Gamma_1\, c_{a_1}(x_1)\,
\left(u_{a_2}(x_2) \,C\Gamma_2 \, s_{a_3}(x_3)\right)\,,
\nn
F_B &=& \delta^{(4)}\Big(x-\sum\limits_{i=1}^3 w_i x_i\Big)
\Phi_B\Big(\sum\limits_{i<j}(x_i-x_j)^2\Big) \,,
\ena
where we follow the notation introduced in Eq.~(\ref{Lagr}).

\begin{figure}[ht]
\begin{center}
\epsfig{figure=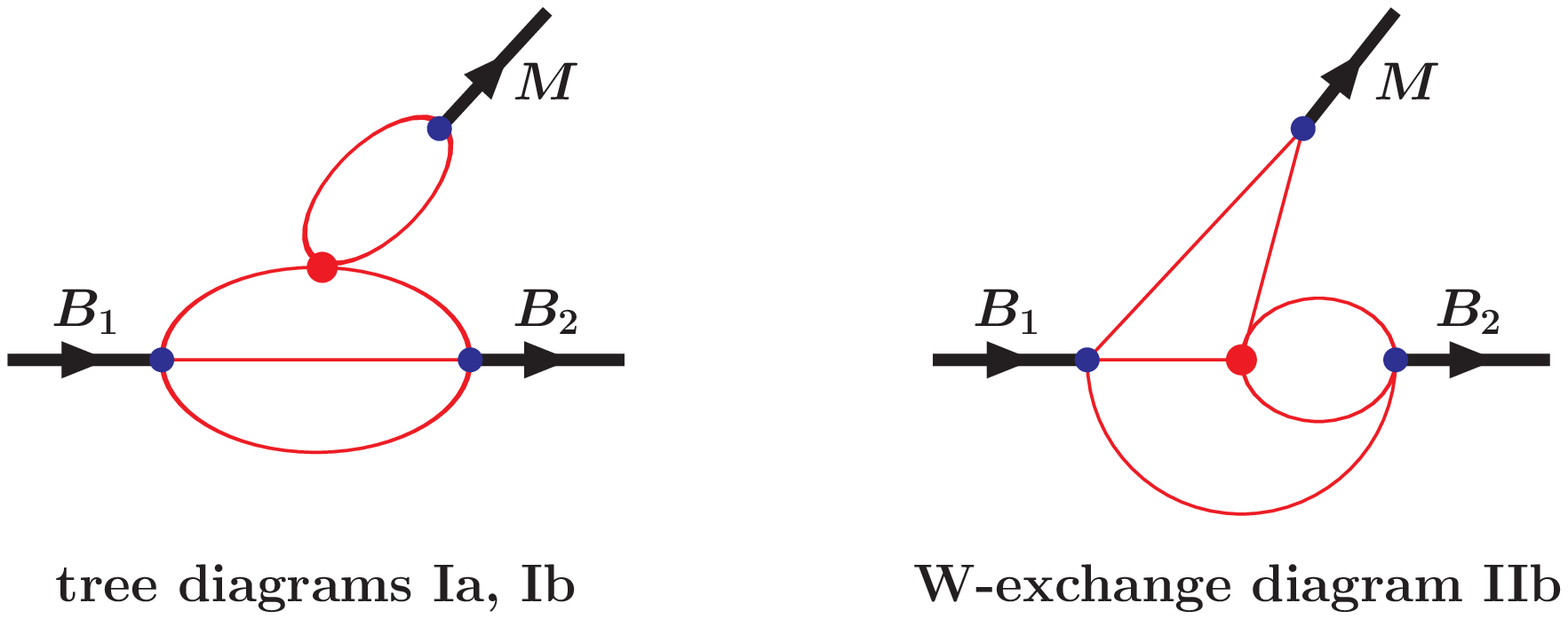,scale=.6}
\caption{Pictorial representations of Eqs.~(\ref{eq:tree})
  and (\ref{eq:W}).}
\label{fig:diag}
\end{center}
\end{figure}

In the following we shall restrict our discussion to the specific 
double charm baryon nonleptonic decay modes
$\Xi_{cc}^{++} \to \Xi_{c}^{(\prime,* )\,+} + \pi^+(\rho^+)$ and
$\Omega_{cc}^{+} \to \Xi_{c}^{(\prime ,*)\,+}+\bar {K}^{(*)0}$ studied in our most
recent paper~\cite{Gutsche:2018msz}. As Table~\ref{tab3}
shows these decays obtain contributions from
the tree diagram and the $W$--exchange topology IIb. The matrix element
can be written as
\bea
<B_2\,M|{\cal H}_{\rm eff}|B_1>
&=&
g_{\rm eff}\,\bar u(p_2)\Big( 12\,C_T\,M_T + 12\, (C_1-C_2)\,M_W \Big)u(p_1).
\label{eq:matr_elem}
\ena
The tree diagram color factor for the neutral $\Omega_{cc}^+$ decays is
given by $C_T=-(C_1+\xi C_2)$ and by $C_T= + (C_2+\xi C_1)$ for the
charged $\Xi_{cc}^+$ decays.
The factor of $\xi=1/N_c$ is set to zero in our numerical calculations
taking $N_c=\infty$.
The overall factor of 12 in Eq.~(\ref{eq:matr_elem}) has its origin in a
combinatorial factor
of 2 and a factor of 6 from the contraction of two Levi-Civita color tensors.
The Feynman diagrams  describing these processes are depicted in
Fig.~\ref{fig:diag}.

The contribution from the tree diagram factorizes into two pieces
according to
\bea
M_T &=& M_T^{(1)}\cdot M_T^{(2)},
\nn
M_T^{(1)} &=&
N_c\,g_{M}\int\!\!\frac{d^4k}{(2\pi)^4i}\widetilde\Phi_{M}(-k^2)\,
\Tr\left[O^\delta_L S_d(k-w_d q) \Gamma_{M} S_{s(u)}(k+w_{s(u)} q)\right]
\nn
M_T^{(2)}&=&
g_{B_1}g_{B_2}\int\!\!\frac{d^4k_1}{(2\pi)^4i}\int\!\!\frac{d^4k_2}{(2\pi)^4i}
\widetilde\Phi_{B_1}\Big(-\vec\Omega_1^{\,2}\Big)
\widetilde\Phi_{B_2}\Big(-\vec\Omega_2^{\,2}\Big)
\nn
&\times&
\Gamma_1 S_c(k_2)\gamma^\mu S_c(k_1-p_1) O_{R\,\delta} S_{u(s)}(k_1-p_2)
\widetilde\Gamma_2 S_{s(u)}(k_1-k_2)\gamma_\mu\gamma_5 \,.
\label{eq:tree}
\ena
Here $\Gamma_1\otimes\widetilde\Gamma_2=+I\otimes\gamma_5$ for
the $\Xi_c^{+}$-baryon and $-\gamma_\nu\gamma_5\otimes\gamma^\nu$
for the $\Xi_c^{\prime\,+}$-baryon.

The coupling constants $g_{M}$, $g_{B_1}$ and  $g_{B_2}$ are determined by the
compositeness condition
as described in our previous papers (for details see,
e.g.~\cite{Gutsche:2017hux,Gutsche:2018utw}).
The Dirac matrix $\Gamma_{M}$ in $M_T^{(1)}$
reads $\gamma_5$ and $\epsilon_V\cdot\gamma$ for the pseudoscalar meson $P$
and for the vector meson $V$, respectively.
The connection of $M_T^{(1)}$ with the leptonic decay constants $f_M=f_P,f_V$
is given by $M_T^{(1)} = -f_P\,q^\delta$ and $+f_V m_V \,\epsilon_V^\delta$ .
The minus sign  in front of $f_P$ appears because
the momentum $q$ flows in the opposite direction from
the decay of $P$-meson.

The calculation of the three-loop $W$--exchange contribution
is much more involved because the matrix element does not factorize.
By using the Fierz transformation 
\[
O_{L/R}^{\alpha_1\alpha_2}O_{R/L}^{\alpha_3\alpha_4}=
2\,(1\pm\gamma_5)^{\alpha_1\alpha_4} (1\mp\gamma_5)^{\alpha_3\alpha_2}
\]
one has
\bea
M_W &=&
g_{B_1}g_{B_2}g_{M}
\int\!\!\frac{d^4k_1}{(2\pi)^4i}
\int\!\!\frac{d^4k_2}{(2\pi)^4i}
\int\!\!\frac{d^4k_3}{(2\pi)^4i}
\widetilde\Phi_{B_1}\Big(-\vec\Omega_1^{\,2}\Big)
\widetilde\Phi_{B_2}\Big(-\vec\Omega_2^{\,2}\Big)
\widetilde\Phi_{M}(-P^2)\,
\nn
&\times&
2\,\Gamma_1\, S_c(k_1)\gamma^\mu S_c(k_2)(1-\gamma_5)S_d(k_2-k_1+p_2)
\Gamma_{M}S_{s(u)}(k_2-k_1+p_1)\gamma_\mu\gamma_5
\nn
&\times&
\Tr\Big[S_{u(s)}(k_3)\widetilde\Gamma_2 S_{s(u)}(k_3-k_1+p_2)(1+\gamma_5)\Big]\,,
\label{eq:W}
\ena
where $\Gamma_1\otimes\widetilde\Gamma_2=I\otimes\gamma_5$ for
$B_2=\Xi^{+}_c$ and $-\gamma_\nu\gamma_5\otimes\gamma^\nu$
for $B_2=\Xi^{\prime\,+}_c$. Here $P = k_2-k_1+w_d\,p_1 + w_u\,p_2$
is the Jacobi momentum in the meson vertex function.

An important simplifying feature of the $W$--exchange contributions is a
straightforward application of
the K\"orner-Pati-Woo (KPW) theorem 
\cite{Korner:1970xq,Pati:1970fg} which states that the contraction of a
flavor-symmetric and flavor-antisymmetric configuration vanishes where
the flavor-antisymmetric configuration follows from the $(V-A)(V-A)$
structure of the current-current interaction. In the present case the KPW
theorem applies to the final state single charm baryon states
$\Xi_{c}^{\prime\,+}$ and $\Xi_{c}^{\ast\,+}$ built from a symmetric light
flavor configuration $(c\{su\})$ which is contracted with the
current-current vertex in the relevant topology Fig. IIb. One thus predicts
that the Fig. IIb contribution to the decay modes
$\Xi_{cc}^{++} \to \Xi_{c}^{(\prime,*)\,+} + \pi^+(\rho^+)$ and
$\Omega_{cc}^{+} \to \Xi_{c}^{(\prime ,*)\,+}+\bar {K}^{(*)0}$ vanishes in the 
$SU(3)$ limit when $m_u=m_s$.

One can trace the working of the KPW theorem directly in our three-loop
calculation. To do this, we change the order
of Dirac matrices in the trace by using the properties of the charge
conjugation matrix. Keeping in mind that $\gamma_5$ does not contribute
to the trace, we have 
\bea
\Tr\Big[S_u(k_3)\gamma_\nu S_s(k_3-k_1+p_2)]
=-\,\Tr\Big[S_s(-k_3+k_1-p_2)\gamma_\nu S_u(-k_3)\Big]\,.
\label{eq:trace}
\ena
We insert Eq.~(\ref{eq:trace}) into  Eq.~(\ref{eq:W}) and shift
the integration variable $k_3\to -k_3+k_1-p_2$. One can check that
$\vec\Omega_2^{\,2}$ goes into itself under this transformation
accompanied by an interchange
of the $u-$ and $s-$ quark masses.
Thus, if $m_u=m_s$ then the amplitude $M_W$ is identical zero which directly
confirms the KPW--theorem. We have checked numerically that the three-loop
integral
vanishes in this limit. $SU(3)$ breaking effects can be
calculated by setting $m_u \neq m_s$. 

Details of the calculation of the loop integrals and the subsequent reduction
of the integration over Fock-Schwinger variables to an integration
over a hypercube may be found in our previous papers (see e.g. the most recent
papers~\cite{Gutsche:2017hux,Gutsche:2018utw}).
Compared to the two-loop calculation
of~\cite{Gutsche:2017hux,Gutsche:2018utw}) we are now dealing with
a three-loop calculation involving  six quark propagators instead of
the four propagators in the two-loop case.
The calculation is quite time-consuming both analytically and numerically.

Next one expands the transition amplitudes in terms of invariant amplitudes.
One has
\bea
&&
<B_2\,P|{\cal H}_{\rm eff}|B_1>
=
g_{\rm eff}\,
\bar u(p_2)\left( A+\gamma_5\,B \right)u(p_1)\,,
\label{eq:ampl-P}
\\
&&
<B_2\,V|{\cal H}_{\rm eff}|B_1>
=
g_{\rm eff}\,
\bar u(p_2)\,\epsilon^\ast_{V \delta}
\left( \gamma^\delta\,V_{\gamma}+p_1^\delta\,V_{p}
+\gamma_5 \gamma^\delta\,V_{5\gamma}+\gamma_5 p_1^\delta\,V_{5p} \right)
u(p_1).
\label{eq:ampl-V}
  \ena
The invariant amplitudes are converted to a set of helicity amplitudes
$H_{\lambda_1\,\lambda_M}$ as described in~\cite{Korner:1992wi}. One has
\bea
 H^V_{\tfrac12\,t} &=& \sqrt{Q_+}\,A\,,\qquad
 H^A_{\tfrac12\,t} = \sqrt{Q_-}\,B\,,
 \nn[-2mm]
 H^V_{\tfrac12\,0} &=&
 + \sqrt{Q_-/q^2}\,\Big( m_+\, V_\gamma + \tfrac12 Q_+\,V_p\Big)\,,
 \,\,\,\,\qquad
 H^V_{\tfrac12\,1}  = -\sqrt{2Q_-}\,V_\gamma\,,
 \nn[-2mm]
 H^A_{\tfrac12\,0} &=&
+ \sqrt{Q_+/q^2}\,\Big( m_-\, V_{5\gamma} + \tfrac12 Q_-\,V_{5p}  \Big)\,,
 \qquad
 H^A_{\tfrac12\,1}  = - \sqrt{2Q_+}\,V_{5\gamma}\,,
 \ena
where $m_\pm=m_1\pm m_2$, $Q_{\pm}=m_\pm^2-q^2$ and
$\mathbf{|p_2|}=\lambda^{1/2}(m_1^2,m_2^2,q^2)/(2m_1)$.
The helicities of the three particles are related by
$\lambda_1 = \lambda_2 - \lambda_M$. We use the notation
$\lambda_P=\lambda_t=0$ for the scalar $(J=0)$ contribution
in order to set the helicity label apart from $\lambda_V=0$
used for the longitudinal component of the $J=1$ vector meson.
The remaining helicity amplitudes can be obtained from the parity relations
$H^V_{-\lambda_2,-\lambda_M} = + H^V_{\lambda_2,\lambda_M}$ and
$H^A_{-\lambda_2,-\lambda_M} = - H^A_{\lambda_2,\lambda_M}$\,.
The helicity amplitudes have the dimension $[m]^3$.
The numerical results on the helicity
amplitudes given in Tables~\ref{tab:OmS}-\ref{tab:XiA} are in units of GeV$^3$.

The two-body decay widths read ($H_{\lambda_2\,\lambda_M}=H^V_{\lambda_2\,\lambda_M}
-H^A_{\lambda_2\,\lambda_M}$)
  \bea
  \Gamma(B_1\to B_2+P) &=&
  \frac{g_{\rm eff}^2}{16\pi}\frac{\mathbf{|p_2|}}{m_1^2}\,
  {\mathcal H}_S\,, \quad
  {\mathcal H}_S =
  \Big|H_{ \tfrac12\,t}\Big|^2 \,+\,
  \Big|H_{-\tfrac12\,t}\Big|^2 \,,
  \label{eq:width_P}
  \\[-2mm]
  \Gamma(B_1\to B_2+V) &=&
  \frac{g_{\rm eff}^2}{16\pi}\frac{\mathbf{|p_2|}}{m_1^2}\,
  {\mathcal H}_V\,, \quad
  {\mathcal H}_V =
  \Big|H_{ \tfrac12\, 0}\Big|^2 \,+\,
  \Big|H_{-\tfrac12\, 0}\Big|^2 \,+\,
  \Big|H_{ \tfrac12\, 1}\Big|^2 \,+\,
  \Big|H_{-\tfrac12\,-1}\Big|^2 \,,
  \label{eq:width_V}
 \ena
where we denote the sum of the squared moduli of the  helicity amplitudes
by ${\cal H}_S$ and ${\cal H}_V$~\cite{Gutsche:2018utw}.

\section{Numerical results}

All model parameters have been fixed in our previous studies except for
the size parameter $\Lambda_{cc}$ of the double charmed baryons. As a first
approximation we equate the size parameter of double charm baryons with
that of single charm baryons, i.e. we take
$\Lambda_{cc}=\Lambda_{c}=0.8675$~GeV where we
adopt the value of $\Lambda_{c}$ from~\cite{Gutsche:2015rrt}
obtained by fitting the magnetic moment of $\Lambda_{c}$ 
to its experimental value. 

Numerical results for the helicity amplitudes and decay widths
are displayed in the Tables~\ref{tab:OmS}-\ref{tab:XiA}.
In this review we concentrate on predictions for the rate values.
In addition to the rate predictions, Tables~\ref{tab:OmS}-\ref{tab:XiA}
contain a
wealth of spin polarization information. For example, for the decay
$\Xi_{cc}^{++}\rightarrow \Xi_{c}^{+}+\pi^{+}$ one finds an asymmetry
parameter of $\alpha = - 2 H^V_{1/2\,0}H^A_{1/2\,0}/(|H^V_{1/2\,0}|^2
+|H^A_{1/2\,0}|^2)=-0.57 $ while~\cite{Sharma:2017txj} predict a value
in the range $\alpha= [-0.86, -1.00]$ depending on their model assumptions.
Note that the $W$--exchange contribution in~\cite{Sharma:2017txj} is purely
$p$--wave, i.e. proportional to $H^A_{1/2\,0}$,
due to the nonrelativistic approximations that they
employ. This is in stark contrast to our relativistic result where the
$s$--wave amplitude dominates in this process,
i.e. $H^V_{1/2\,0}/H^A_{1/2\,0}=3.3$. Both model calculations agree on a very
substantial destructive interference of the tree and $W$--exchange
contributions.

Our results highlight the importance of the KPW theorem for the nonleptonic
decays when the final state involves a $\Xi^{\prime+}$ baryon containing a
symmetric
$\{su\}$ diquark. Tables~\ref{tab:OmS}-\ref{tab:XiA} show that the relevant
$W$--exchange contributions are nonzero  but
are strongly suppressed. Nonzero values result
from $SU(3)$ breaking effects which are accounted for in our approach.
Take for example the decay $\Xi_{cc}^{++} \to \Xi_c^{'+} +\pi^+$. When
compared to the tree contribution the $SU(3)$ breaking effects amount
to $\sim(2-4)\,\%$. While the consequences of the KPW theorem for the
$W$--exchange contribution are also incorporated in the pole model approach
of~\cite{Sharma:2017txj} they are not included in the final-state interaction
approach of~\cite{Jiang:2018oak}.

In Table~\ref{tab:comparison} we compare our rate results with the
results of some other
approaches~\cite{Dhir:2018twm,Sharma:2017txj,Jiang:2018oak,Wang:2017mqp,Yu:2017zst,Kiselev:2001fw}.
For convenience purposes, we put the columns in orders
corresponding to the number of calculated modes.
We put Jiang et al. before  Wang et al. according to  alphabetical 
order. 
The rates calculated in~\cite{Wang:2017mqp} include tree graph contributions only.
There is a wide spread in the rate values predicted by the various model calculations. All
calculations approximately agree on the rate of the decay
$\Xi_{cc}^{++} \to \Xi_c^{'+} +\rho^+$ which is predicted
to have a large branching
ratio of $\sim 16\, \%$. In our calculation this mode is predicted to have
by far the largest branching ratio of the decays analyzed in this paper.
As concerns the decay $\Xi_{cc}^{++}\rightarrow \Xi_{c}^{+}+\pi^{+}$ discovered
by the LHCb Collaboration~\cite{Aaij:2018gfl} we find a branching ratio of
${\cal B}(\Xi_{cc}^{++}\rightarrow \Xi_{c}^{+}\pi^{+})=0.70 \,\%$
using the central value of the life time measurement in~\cite{Aaij:2018wzf}.
The small value of the branching ratio results from a substantial
cancellation of the tree and $W$--exchange contributions. 
The branching ratio is somewhat smaller than the branching ratio
${\cal B }(\Xi_{cc}^{++} \to \Sigma_c^{++} + \bar K^0)=1.28\,\%$ calculated
in~\cite{Gutsche:2017hux}. We think that the latter mode is more dominant 
in comparison with $\Xi_{cc}^{++}\rightarrow \Xi_{c}^{+}\pi^{+}$. 
We predict a branching ratio considerably smaller than the range of
branching fractions $(6.66 - 15.79)\,\%$ calculated in~\cite{Sharma:2017txj}. 
In our opinion the calculations done in Ref.~\cite{Sharma:2017txj} 
involve generous approximations for the errors which are hard to quantify.

An important issue is the accuracy of our results.
The only free parameter in our approach is the size parameter
$\Lambda_{cc}$ of the double heavy baryons for which 
we have chosen $\Lambda_{cc}=0.8675$~GeV in Tables~\ref{tab:OmS}-\ref{tab:XiA}.
In order to estimate the uncertaintity caused by the choice of the
size parameter we allow the size parameter to vary from 0.6 to 1.135~GeV.
We evaluate the mean $\bar\Gamma = \sum\Gamma_i/N$ and the mean 
square deviation
$\sigma^2=\sum (\Gamma_i-\bar\Gamma)^2/N$. The results for $N=5$
are shown in Table~\ref{tab:errors}. The rate errors amount to $6 - 15 \%$.
Since the dependence of the rates on $\Lambda_{cc}$ is nonlinear the central
values of the rates in Table~\ref{tab:errors} do not agree
with the rate values in~Tables~\ref{tab:OmS}-\ref{tab:XiA}.

\begin{table}[!htbp]
\begin{tabular}{cc}
    \begin{minipage}{.45\linewidth}
   \centering
   \caption{
      Decays 
$\Omega^+_{cc}\to\Xi^{\prime\,+}_{c} + \bar K^0(\bar K^{\ast\,0})$}
    \label{tab:OmS}
    \vskip 1 mm
     \def\arraystretch{.9}
\begin{tabular}{|cccc|}
\hline
  Helicity        & Tree diagram &  $W$ diagram & total  \\
\hline
$ H^V_{\tfrac12\,t} $ &  $0.20$     & $-0.01$   & $0.19$  \\
$ H^A_{\tfrac12\,t} $ &  $0.25$     & $-0.01$   & $0.24$
\\[1.1ex]
\hline
\multicolumn{4}{|c|}
            {$\Gamma(\Omega^+_{cc}\to\Xi^{\prime\,+}_{c}+ \bar K^0)                                                
              = 0.15\cdot 10^{-13}\,\text{GeV}\ $} \\
\hline
$ H^V_{\tfrac12\,0} $ &  $-0.25$    & $0.04 \times 10^{-1}$    &  $-0.25$ \\
$ H^A_{\tfrac12\,0} $ &  $-0.50$    & $ 0.01$    &  $-0.49$ \\
$ H^V_{\tfrac12\,1} $ &  $ 0.27$    & $-0.01$    &  $ 0.26$ \\
$ H^A_{\tfrac12\,1} $ &  $ 0.56$    & $0.04 \times 10^{-2}$    &  $ 0.56$
\\[1.1ex]
\hline
\multicolumn{4}{|c|}
            {$\Gamma(\Omega^+_{cc}\to\Xi^{\prime\,+}_{c}+ \bar K^{\ast\,0})                                        
              = 0.74 \cdot 10^{-13}\,\text{GeV}\ $} \\
\hline
\end{tabular}
\end{minipage} &

    \begin{minipage}{.45\linewidth}

   \centering
   \caption{
      Decays $\Omega^+_{cc}\to\Xi^{+}_{c} + \bar K^0(\bar K^{\ast\,0})$ }
    \label{tab:OmA}
    \vskip 1 mm
     \def\arraystretch{.9}
\begin{tabular}{|cccc|}
\hline
  Helicity        & Tree diagram &  $W$ diagram & total  \\
\hline
$ H^V_{\tfrac12\,t} $ &  $-0.35$     & $1.06$   & $0.71$  \\
$ H^A_{\tfrac12\,t} $ &  $-0.10$     & $0.31$    & $0.21$
\\[1.1ex]
\hline
\multicolumn{4}{|c|}
            {$\Gamma(\Omega^+_{cc}\to\Xi^{+}_{c}+ \bar K^0)                                                   
              = 0.95 \cdot 10^{-13}\,\text{GeV}\ $} \\
\hline
$ H^V_{\tfrac12\,0} $ &  $ 0.50$   & $-0.69$   &  $-0.19$ \\
$ H^A_{\tfrac12\,0} $ &  $ 0.18$   & $-0.45$   &  $-0.27$ \\
$ H^V_{\tfrac12\,1} $ &  $-0.11$   & $-0.24$   &  $-0.35$ \\
$ H^A_{\tfrac12\,1} $ &  $-0.18$  & $ 0.66$    &  $ 0.48$
\\[1.1ex]
\hline
\multicolumn{4}{|c|}
            {$\Gamma(\Omega^+_{cc}\to\Xi^{+}_{c}+ \bar K^{\ast\,0})                                           
                = 0.62\cdot 10^{-13}\,\text{GeV}\ $} \\
\hline
\end{tabular}
\end{minipage}
\end{tabular}

\begin{tabular}{cc}
    \begin{minipage}{.45\linewidth}
   \centering
   \caption{
      Decays $\Xi^{++}_{cc}\to\Xi^{\prime\,+}_{c} + \pi^+(\rho^+)$ }
    \label{tab:XiS}
    \vskip 1 mm
     \def\arraystretch{.9}
\begin{tabular}{|cccc|}
\hline
  Helicity      & Tree diagram &  $W$ diagram & total  \\
\hline
$ H^V_{\tfrac12\,t} $ &  $-0.38$     & $-0.01$ & $-0.39$  \\
$ H^A_{\tfrac12\,t} $ &  $-0.55$     & $-0.02$ & $-0.57$
\\[1.1ex]
\hline
\multicolumn{4}{|c|}
            {$\Gamma(\Xi^{++}_{cc}\to\Xi^{\prime\,+}_{c} + \pi^+)                                                  
              = 0.82\cdot 10^{-13}\,\text{GeV}\ $} \\
\hline
$ H^V_{\tfrac12\,0} $ &  $ 0.60$   & $0.04 \times 10^{-1}$   &  $0.61$ \\
$ H^A_{\tfrac12\,0} $ &  $ 1.20$   & $0.01$   &  $1.21 $ \\
$ H^V_{\tfrac12\,1} $ &  $-0.49$   & $-0.01$  &  $-0.50$ \\
$ H^A_{\tfrac12\,1} $ &  $-1.27$   & $0.01 \times 10^{-1}$ &  $-1.27$
\\[1.1ex]
\hline
\multicolumn{4}{|c|}
            {$\Gamma(\Xi^{++}_{cc}\to\Xi^{\prime\,+}_{c} + \rho^+)                                            
               = 4.27\cdot 10^{-13}\,\text{GeV}\ $} \\
\hline
\end{tabular}
\end{minipage} &

    \begin{minipage}{.45\linewidth}

   \centering
   \caption{
      Decays $\Xi^{++}_{cc}\to\Xi^{+}_{c} + \pi^+(\rho^+)$}
    \label{tab:XiA}
     \vskip 1 mm
    \def\arraystretch{.9}
\begin{tabular}{|cccc|}
\hline
  Helicity         & Tree diagram &  $W$ diagram & total  \\
\hline
$ H^V_{\tfrac12\,t} $ &  $-0.70$     & $0.99$ & $0.29$  \\
$ H^A_{\tfrac12\,t} $ &  $-0.21$     & $0.30$ & $0.09$
\\[1.1ex]
\hline
\multicolumn{4}{|c|}
            {$\Gamma(\Xi^{++}_{cc}\to\Xi^{+}_{c} + \pi^+)                                                     
              = 0.18\cdot 10^{-13}\,\text{GeV}\ $} \\
\hline
$ H^V_{\tfrac12\,0} $ &  $ 1.17$    & $-0.70$   &  $ 0.47$  \\
$ H^A_{\tfrac12\,0} $ &  $ 0.45$    & $-0.44$   &  $ 0.003$ \\
$ H^V_{\tfrac12\,1} $ &  $-0.20$    & $-0.23$   &  $-0.43$  \\
$ H^A_{\tfrac12\,1} $ &  $-0.41$    & $ 0.62$   &  $ 0.21$
\\[1.1ex]
\hline
\multicolumn{4}{|c|}
            {$\Gamma(\Xi^{++}_{cc}\to\Xi^{+}_{c} + \rho^+)          
              = 0.63\cdot 10^{-13}\,\text{GeV}\ $} \\
\hline
\end{tabular}
\end{minipage}
\end{tabular}
\end{table}
\begin{table}[!htbp]
\vskip 1mm   
   \centering
   \caption{Comparison with other approaches. Abbreviation: M=NRQM, T=HQET}
    \label{tab:comparison}
     \def\arraystretch{1.2}
\begin{tabular}{|l|c|c|c|c|c|c|}
\hline
\qquad Mode &  \multicolumn{6}{|c|}{Width (in $10^{-13}$~GeV)} \\
\cline{2-7}
& \qquad  GIKLT~\cite{Gutsche:2017hux,Gutsche:2018msz}
\qquad\qquad & DS~\cite{Dhir:2018twm,Sharma:2017txj}
& JHL~\cite{Jiang:2018oak} & WYZ~\cite{Wang:2017mqp}
& YJLLWZ~\cite{Yu:2017zst}    &  KL~\cite{Kiselev:2001fw}
\\
\hline   
$\Xi^{++}_{cc}\to\Sigma^{++}_{c} + \bar K^0$  & 0.33   & & & & &
\\
\hline   
$\Xi^{++}_{cc}\to\Sigma^{++}_{c} + \bar K^{\ast\,0}$  & 1.38   & & & & &
\\
\hline
$\Omega^+_{cc}\to\Xi^{\prime\,+}_{c} + \bar K^0$  & 0.15   & 0.31 (M)  & & & &
\\
                                                  &        & 0.59 (T)  & & & &
\\
\hline
$ \Omega^+_{cc}\to\Xi^{+}_{c}+ \bar K^0 $         & 0.95  & 0.68 (M) &  & & &
\\
                                                  &        & 1.08 (T) &  & & &
\\
\hline
$\Omega^+_{cc}\to\Xi^{\prime\,+}_{c} + \bar K^{\ast\,0}$ & 0.74
& & $2.64^{+2.72}_{-1.79}$ &  & &
\\
\hline
$\Omega^+_{cc}\to\Xi^{+}_{c}+ \bar K^{\ast\,0}$& 0.62 & & $1.38^{+1.49}_{-0.95}$
&  & &
\\
\hline
$\Xi^{++}_{cc}\to\Xi^{\prime\,+}_{c} + \pi^+$  & 0.82 &1.40 (M) &   & 1.10 & &
\\
                                               & & 1.93 (T)&      &        & &
\\
\hline
$\Xi^{++}_{cc}\to\Xi^{+}_{c} + \pi^+$          & 0.18 &1.71 (M) &   & 1.57
& 1.58 & 2.25
\\
                                               & & 2.39 (T)&   & & &
\\
\hline
$\Xi^{++}_{cc}\to\Xi^{\prime\,+}_{c} + \rho^+$ & 4.27 & & $4.25^{+0.32}_{-0.19}$
& 4.12 & 3.82 &
\\
\hline
$\Xi^{++}_{cc}\to\Xi^{+}_{c} + \rho^+$         & 0.63 & & $4.11^{+1.37}_{-0.86}$
& 3.03 & 2.76 & 6.70
\\
\hline
\end{tabular}
\end{table}
\begin{table}[!htbp]
\vskip 1mm 
\centering   \caption{Estimating uncertainties in the decay widths.}
    \label{tab:errors}
     \def\arraystretch{1.2}
\begin{tabular}{|lc|}
  \hline
\qquad   Mode \qquad  & \qquad Width (in $10^{-13}$~GeV) \qquad \\
  \hline
  $\Omega^+_{cc}\to\Xi^{\prime\,+}_{c} + \bar K^0$ \qquad &
\qquad $0.14 \pm  0.01$ \qquad  
\\  
$\Omega^+_{cc}\to\Xi^{\prime\,+}_{c} + \bar K^{\ast\,0}$ \qquad &
\qquad $0.72 \pm  0.06$ \qquad 
\\
\hline
$\Omega^+_{cc}\to\Xi^{+}_{c} + \bar K^0$ \qquad   &
\qquad $0.87 \pm 0.13$ \qquad 
\\
$\Omega^+_{cc}\to\Xi^{+}_{c} + \bar K^{\ast\,0}$ \qquad &
\qquad $0.58 \pm 0.07$ \qquad 
\\
\hline
$\Xi^{++}_{cc}\to\Xi^{\prime\,+}_{c} + \pi^+$ \qquad  &
\qquad $0.77 \pm 0.05$ \qquad 
\\
$\Xi^{++}_{cc}\to\Xi^{\prime\,+}_{c} + \rho^+$ \qquad &
\qquad $4.08 \pm 0.29$ \qquad 
\\
\hline
$\Xi^{++}_{cc}\to\Xi^{+}_{c} + \pi^+$ \qquad &
\qquad $0.16 \pm 0.02$\qquad 
\\
$\Xi^{++}_{cc}\to\Xi^{+}_{c} + \rho^+$ \qquad &
\qquad $0.59 \pm 0.04$ \qquad
\\
\hline
\end{tabular} 
\end{table}

\section{Summary and outlook}

We have proposed a calculational technique which allows one to evaluate the
$W$-exchange graphs in nonleptonic decays of heavy and light baryons.
In this review we have concentrated on the description of Cabibbo-favored
nonleptonic two-body decays of the double charm
ground state baryons $\Xi_{cc}^{++}$, $\Xi_{cc}^{+}$, and $\Omega_{cc}^+$
where we have limited our analysis to the $1/2^+ \to 1/2^+ + P(V)$ decay
channels. It would be straightforward to also include the
$1/2^+ \to 3/2^+ + P(V)$ nonleptonic decays not discussed in this review.
Also the study could be extended to the description of singly and doubly
suppressed Cabibbo decays not only for double charm baryon decays but also for
single charm baryon decays.
In the future we plan to extend our predictions for other modes of
nonleptonic decays 
of double and single heavy baryons taking into account $W$-exchange
contributions. 
One can see from our analysis that the $W$--exchange contributions are
generally not suppressed. 
Moreover, one can identify decay modes that are contributed to only by
$W$-exchange graphs. 
A typical example is the doubly Cabibbo-suppressed decay $\Xi_{c}^{+}\to p\phi$
recently observed by
the LHCb Collaboration~\cite{Aaij:2019kss} which is induced by the quark
level transition $(c\to d;\,s\to u)$. In our classification it is contributed
to by the topology diagram IIb. 

\funding{This work was funded by 
the Carl Zeiss Foundation under Project ``Kepler Center f\"ur Astro-
und Teilchenphysik: Hochsensitive Nachweistechnik zur Erforschung des
unsichtbaren Universums (Gz: 0653-2.8/581/2)'', by CONICYT (Chile) PIA/Basal FB0821.  
M.A.I.\ acknowledges the support from the PRISMA$^+$ Cluster of Excellence (Mainz Uni.). 
M.A.I. and J.G.K. acknowledge the support of a Heisenberg-Landau grant.}

\conflictsofinterest{The authors declare no conflict of interest}

\reftitle{References}

\end{document}